\documentclass[a4paper,11pt]{article}
\usepackage{pos}
\usepackage{slashed}

\title{Heavy quark drag coefficient using phonon-heavy quark EFT}

\author*[a]{Andreas Kirchner}
\author[a]{Noah Koliadko}
\author[a]{Berndt Mueller}
\author[a]{Jyotirmoy Roy}

\affiliation[]{Department of Physics, Duke University, Durham, NC 27708, USA }

\emailAdd{andreas.kirchner@duke.edu}
\emailAdd{noah.koliadko@duke.edu}
\emailAdd{jyotirmoyr@smu.edu}
\emailAdd{berndt.mueller@duke.edu}

\abstract{The description of heavy quarks inside the QGP medium, especially at low momenta, remains challenging from first principles due to the strong coupling and large gluon occupation numbers. An alternative description in terms of perturbative degrees of freedom treats the interactions of heavy quarks with the collective excitations of the medium in the form of phonons. This description is derived as an effective field theory (EFT) and therefore allows for a naturally occurring power counting of the allowed interactions. This EFT has already been established in our previous work \cite{Kirchner:2025xdb}, where we derived the phonon-heavy quark interactions and calculated their thermal scattering cross section.
In this work, we use these results, together with the heavy quark expansion, to calculate the heavy quark drag coefficient. Subsequently, we compare our results with the drag coefficient obtained from AdS/CFT calculations to determine the previously unknown Wilson coefficients for the quark-phonon coupling. Additionally, we determine the same Wilson coefficients by matching the EFT result for the drag force at low momenta with the QCD result at high momenta. Finally, we will compare our results with experimentally determined heavy quark diffusion coefficients.}

\FullConference{
13th International Conference on Hard and Electromagnetic Probes of High-Energy Nuclear Collisions
}

\begin{document}
\maketitle

\section{Introduction}

Highly energetic probes traversing a medium, such as high-energy quarks in the quark--gluon plasma (QGP), deposit part of their energy and momentum in the medium in the form of a wake or a Mach cone through interactions with it. The perturbations created in this process consist primarily of phonons and shear modes, i.e., the long-range effective degrees of freedom of the medium. A consistent description of these phenomena therefore requires an effective field theory for the interactions between the heavy quark and the collective degrees of freedom of the medium.

We will begin by giving a short overview of the construction of the EFT and then discuss the interactions between the collective degrees of freedom of the fluid and the heavy quark (HQ). Next, we will study the Quark--phonon interactions in greater detail and use these to describe the thermalization process of the heavy quark and the wake formation in the medium. A more detailed description of the results presented here can be found in \cite{Kirchner:2025xdb}.

\section{Construction of the effective field theory}
We begin the construction of the EFT with the fluid degrees of freedom, described by the comoving coordinates of the fluid volume elements $\phi_I(t,\Vec{x})$, with $I=1,2,3$, which track the motion of the individual fluid cells during the evolution of the fluid. From these label fields, we construct the current
\begin{equation}
J^{\mu} = \frac{1}{3!}\varepsilon^{\mu \nu \rho \sigma} \varepsilon_{I J K} \partial_{\nu} \phi^{I} \partial_{\rho} \phi^{J} \partial_{\sigma} \phi^{K}.
\end{equation}
Using the symmetries of the theory \cite{Kirchner:2025xdb}, we construct the fluid Lagrangian $\mathcal{L}=F(X=J^\mu J_\mu)$. The function $F(X)$ can be understood as the equation of state of the fluid, allowing us to calculate thermodynamic variables such as the energy density $\epsilon=-F(X)$ which becomes evident when calculating the energy--momentum tensor of the theory  \cite{Endlich:2010hf, Dubovsky:2011sj}. The coupling to QCD matter is then achieved by elevating the coefficients and couplings of the QCD Lagrangian to functions of the variable $X$. The terms relevant to this work are
\begin{equation}
    \mathcal{L} \supset F(X) + Z_\psi(X) \Bar{\psi}\frac{i}{2} \overleftrightarrow{\slashed{D}} \psi - M(X) \Bar{\psi} \psi
\end{equation}
where $\psi$ is the heavy quark field. The full Lagrangian is given in \cite{Kirchner:2025xdb}.

Next, we introduce small fluctuations around the equilibrium configuration of a fluid at rest, $\phi^I=x^I + \lambda \pi^I(t,\Vec{x})$. Applying this expansion to the terms of the Lagrangian, we find the quadratic part of the fluid Lagrangian to be that of a wave traveling at the speed of sound $c_s$ of the medium. The higher-order terms of this expansion give the interaction terms between the displacement fields\footnote{The phonons interacting with the heavy quarks are the longitudinal modes of the displacement fields.} $\pi^I$. Canonically normalizing our theory by a field redefinition $\pi^I \to \sqrt{-F'(1)} \pi^I$ allows us to introduce the UV scale of our theory $\Lambda_h=\sqrt[4]{-F'(1)}\approx 5T$, which suppresses higher-order interactions. The estimate of five times the temperature $T$ follows from kinematic considerations \cite{Kirchner:2025xdb}. Applying the same expansion and rescaling to the mass term of the heavy quark leads to $M(X)\Bar{\psi} \psi =  \Bar{\psi} \psi (M(1)+ \frac{\Tilde{M}'(1)}{\Lambda_h} \nabla \cdot \Vec{\pi} + \ldots)$ with higher-order interactions between the heavy quark and multiple phonons being more suppressed by the UV scale.

\section{Quark--phonon interactions}

\begin{figure}[t]
	\centering
	\includegraphics[width=0.7\textwidth]{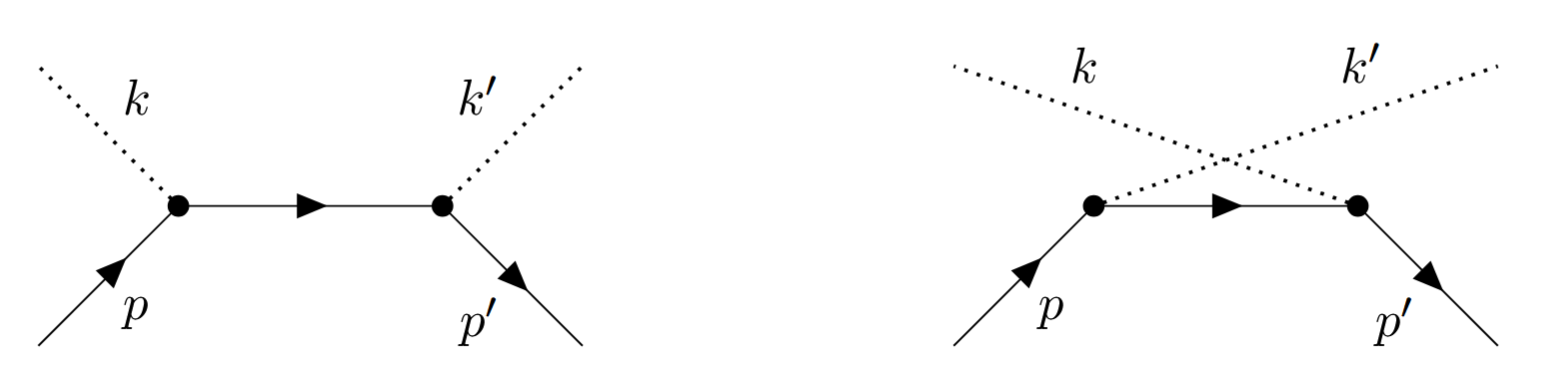}
	\caption{Leading order ($\Lambda_h^{-2}$) $2 \to 2 $ heavy quark-phonon scattering diagrams}\label{Fig:Feynman}
\end{figure}

We now examine the interactions between heavy quarks and phonons through their leading-order scattering cross section. The leading-order $ 2 \to 2 $ diagram is suppressed by $\Lambda_h^{-2}$ with the relevant diagrams being shown in Fig.~\ref{Fig:Feynman}.
Since the phonons are space-like particles, the heavy quark propagator can go on shell, requiring the use of a quasi particle width for the heavy quark. The heavy quark obtains this width through the absorption of thermal phonons and the emission of phonons if the quark's velocity is larger than the speed of sound of the medium (i.e. a Mach cone). Using the quark width, which is finite at all velocities, to calculate the cross section, we find two distinct and interesting features. We find that the cross section is very large in the collinear limit due to the increased interaction time between the heavy quark and the phonon. The cross section shows a second peak, whose position depends on the speed of sound and the momentum of the phonon through the Mach condition (see \cite{Kirchner:2025xdb} for details).

\section{Quark energy loss}

\begin{figure}[htbp]
	\centering
	\includegraphics[height=4.6cm]{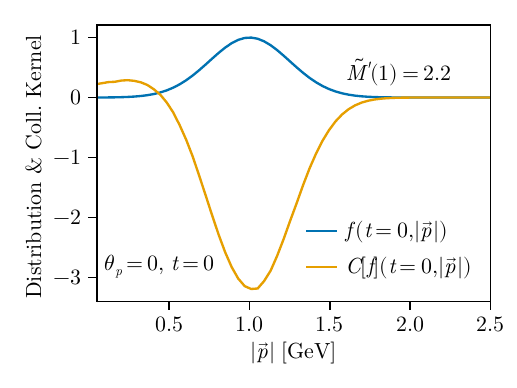}%
	\hspace{1cm}
	\includegraphics[height=4.6cm]{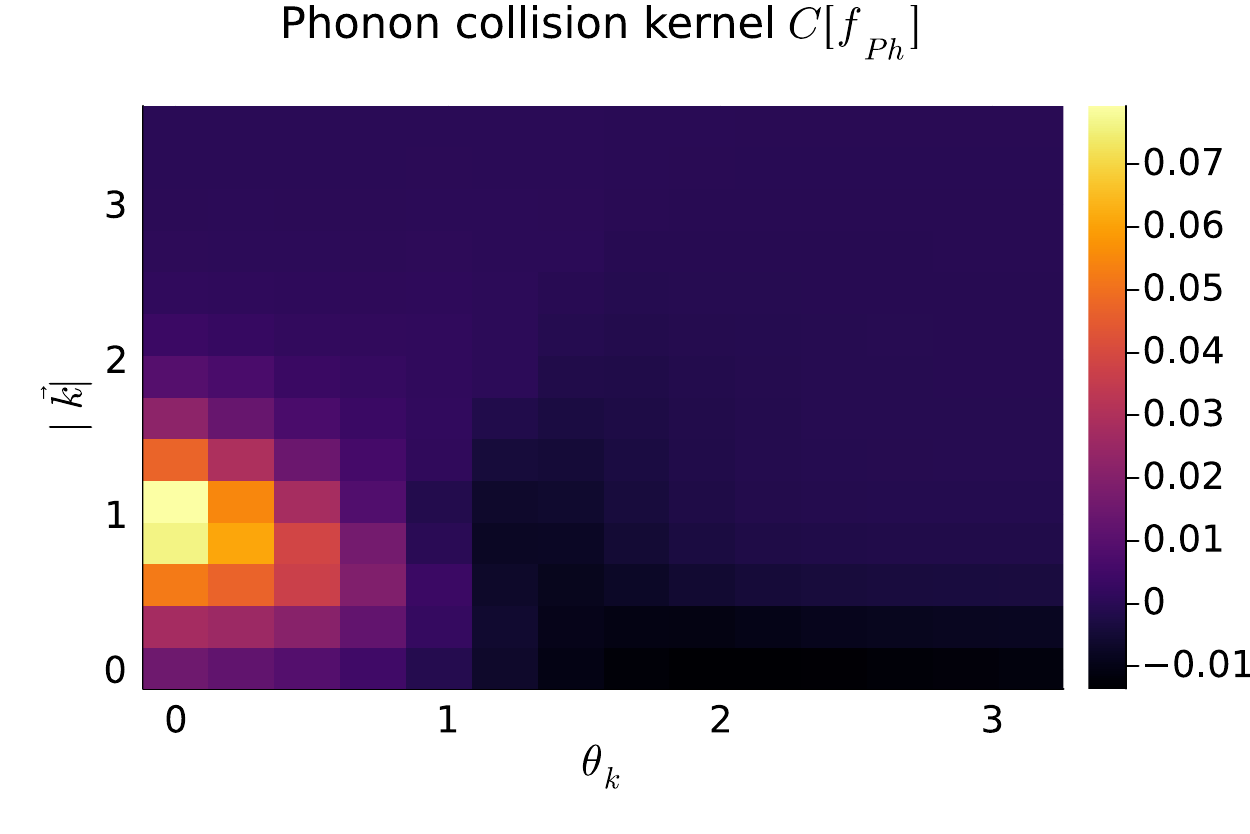}
	\caption{Heavy quark distribution and heavy quark collision kernel for heavy quark polar angle $\theta_p=0$ (left panel) and phonon collision kernel from the same heavy quark distribution function (right panel).}\label{Fig:CollKernel}
\end{figure}

In order to connect the scattering matrix element derived in the last section with experimentally observable quantities, we will study the heavy quark energy loss inside the plasma through interactions between the heavy quark and the phonons of the plasma using the Boltzmann equation. We will assume that both the medium and the heavy quark distribution are spatially homogeneous, allowing us to write the Boltzmann equation as
\begin{equation}
    \partial_t f (t,p) = C[f](t,p),    
\end{equation}
where $f(t,p)$ is the HQ distribution at time $t$ and momentum $p$, and $C[f](t,p)$ is the collision kernel evaluated using the matrix element given in Fig.~\ref{Fig:Feynman}. The result is shown in the left panel of Fig.~\ref{Fig:CollKernel}, with the initial HQ distribution chosen to be a Gaussian centered at $\Vec{p}_0=(0,0,1) \; \mathrm{GeV}$.

\begin{figure}[htbp]
	\centering
	\includegraphics[height=4.6cm]{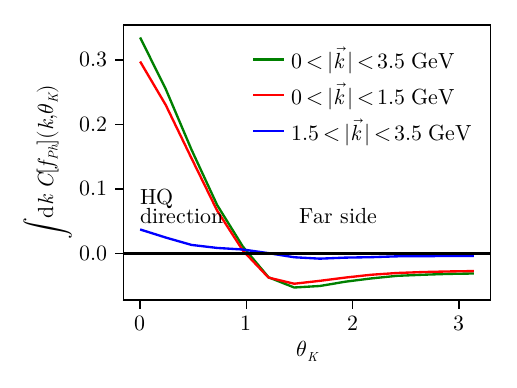}%
	\hspace{1.5cm}
	\includegraphics[height=4.6cm]{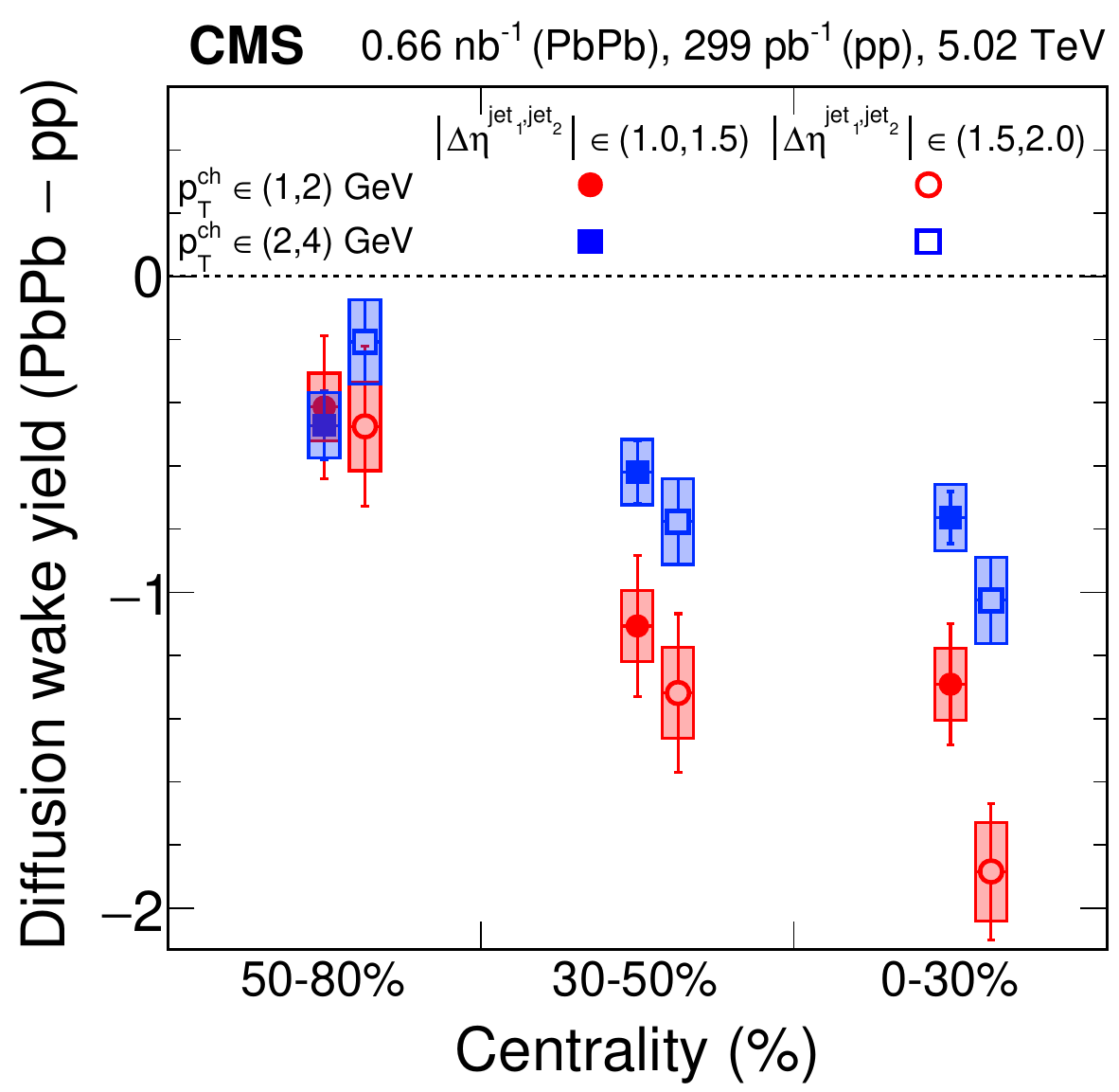}
	\caption{Heavy quark wake signal in the phonon distribution as a function of the polar angle (left) and experimentally observed diffusion wake yield \cite{CMS:2026mur} (right). }\label{Fig:ExpWake}
\end{figure}

Solving the Boltzmann equation, we find that the heavy quark distribution thermalizes and isotropizes (see \cite{Kirchner:2026fxu} for a plot of the evolution) through interactions with the medium and eventually depends only on the magnitude of the heavy quark momentum, becoming independent of its polar angle, displaying the characteristic memory loss found in thermalization processes.
Finally, we also study changes in the medium by calculating the collision kernel for the phonon distribution. The result is shown in the right panel of Fig.~\ref{Fig:CollKernel}. We find that the phonon distribution is enhanced in the direction of motion of the heavy quark and depleted in the opposite direction. This enhancement and depletion is characteristic of a diffusion wake that forms as the heavy quark traverses the medium.
A comparison between the phonon wake signal (left panel) and the experimentally observed diffusion wake \cite{CMS:2026mur} (right panel) is displayed in Fig.~\ref{Fig:ExpWake}.
The full wake signal in the phonon distribution as a function of the angle (green line, left panel) shows the typical enhancement of phonons in the direction of motion of the HQ and depletion on the far side, characteristic of a diffusion wake. When comparing its low-momentum (red line, left panel) and high-momentum (blue line, left panel) parts with the corresponding low (red dots, right panel) and high (blue dots, right panel) measurements, we find a similar trend of a significantly stronger wake signal in the low-momentum contributions.

\section{Conclusion}

In this work, we have developed an EFT describing the interactions between the phonons of the QGP and heavy quarks with a natural power counting. Within this theory, we derived the Feynman rules used for the calculation of the quasi particle width of the heavy quark and its scattering cross section with phonons. Using this scattering cross section, we have calculated the collision kernel of the Boltzmann equation, which in turn was used to study the thermalization dynamics of the heavy quarks in the QGP medium. Finally, we demonstrated the formation of a wake in the medium by calculating the phonon collision kernel and found trends in the wake strength similar to those observed experimentally. In future work, we want to extend this theory to include viscosities and study their effects on the thermalization dynamics of the heavy quarks.


\begin{thebibliography}{99}


\bibitem{Kirchner:2025xdb}
A.~Kirchner, B.~M{\"u}ller, J.~Roy and C.~Sirimanna,
JHEP \textbf{08} (2026), 172
doi:10.1007/JHEP08(2026)172
[arXiv:2510.13942 [hep-ph]].

\bibitem{CMS:2026mur}
A.~Hayrapetyan \textit{et al.} [CMS],
Phys. Rev. Lett. \textbf{137} (2026) no.7, 071902
doi:10.1103/g49y-8cjl
[arXiv:2602.19431 [nucl-ex]].


	\bibitem{Endlich:2010hf}
	S.~Endlich, A.~Nicolis, R.~Rattazzi and J.~Wang,
	JHEP \textbf{04} (2011), 102
	doi:10.1007/JHEP04(2011)102
	[arXiv:1011.6396 [hep-th]].
	
	\bibitem{Dubovsky:2011sj}
	S.~Dubovsky, L.~Hui, A.~Nicolis and D.~T.~Son,
	Phys. Rev. D \textbf{85} (2012), 085029
	doi:10.1103/PhysRevD.85.085029
	[arXiv:1107.0731 [hep-th]].

\bibitem{Kirchner:2026fxu}
A.~Kirchner, B.~Mueller, J.~Roy and S.~Waeber,
[arXiv:2611.01309 [hep-ph]].
\end{thebibliography}
\end{document}